\documentclass{aa}  

\usepackage{graphicx}
\usepackage{txfonts}
\usepackage{caption}
\usepackage{subcaption}
\begin{document}

	\title{Multimessenger signal from phase transition of neutron star to quark star}
	\authorrunning{D.Kuzur et. al}
	
	\author{Debojoti Kuzur
		\inst{1}, Ritam Mallick
		\fnmsep\thanks{}\inst{1}, Prasad R
		\inst{1} \and Shailendra Singh \inst{1}
	}
	
	\institute{Indian Institute of Science Education and Research, Bhopal\\
		\email{mallick@iiserb.ac.in}
	}
	
	
	
	\abstract
	{}
	{To study the multimessenger nature of the signal that can result from the phase transition of a neutron star to a quark star and their corresponding astrophysical observations.}
	{The phase transition process is initiated by the abrupt pressure and density changes at the star center, giving rise to a shock which deconfines matter followed by a weak front converting excess down to strange quarks to attain absolute stability. This process's effects are investigated by understanding how the energy escapes from the star in the form of neutrino-antineutrino annihilation. For such annihilation process, the corresponding energy deposition rate is calculated. Structural changes due to the energy loss have been investigated in the likes of misalignment angle evolution of the star and its astrophysical observation through gravitational waves.}
	{The energy and time signature for the neutrino-antineutrino annihilation is compared with the observed isotropic energy for a short gamma-ray burst. The misalignment angle evolves to align the star's tilt axis, which can lead to the sudden increase or decrease of radio intensity from the pulsar. The corresponding gravitational wave emission, both continuous and burst, all lead towards multimessenger signals coming from the phase transition.}
	{}
	
	
	\maketitle
	%
	
	\section{Introduction}
	The search for the properties of matter at high density/temperature is being investigated for quite some time. The property of matter at the high energy zero chemical potential (zero density) is carried out by the heavy-ion collider experiments. It has been established that the quark-gluon plasma state exists at high-temperature \citep{shuryak,gyulassy,andronic}. However, there are still no earth-based experiments in the high-density sector, and to know the properties of matter at such densities, we still rely on astrophysical observation coming from neutron stars (NS). NS  serves as a natural laboratory for testing matter properties at large densities because of their compactness. 
	
	However, the task is not that easy as NS cores' direct observation cannot be carried out. Observation of NS comes only from their surface and what physicists do is model the star from the core to the surface and then match their observation signatures. Again, modeling the matter properties at high density is still a challenge as ab initio calculation fails \citep{lattimer,forcrand}. However, the situation is improving day by day with new observations coming from different astrophysical detectors. The accurate measurement of massive pulsars in the last decade, on the one hand, \citep{demorest,antonidis,cromartie} and the gravitational detection of binary NS merger (BNSM) on the other \citep{abbott,abbott1} has set a stringent constraint on the equation of state (EoS), which describes matter at such high densities. 
	
	One of the debates that have come to the forefront is whether quark matter (QM) exists at NS's cores\citep{hinderer,read,pozzo,agathos,chatzi,annala,most,rana,zhang,bauswein}. Absolute stable QM is expected to have strange quarks and deconfinement transition, and weak interaction at NS cores is expected to produce stable strange quark matter (SQM)  \citep{witten,abhijit,prasad,prasad1}. The deconfinement transition most probably happens during the formation of NSs after the supernova or after BNSM. However, it can even occur in cold NSs initiated by sudden density rise at the star core \citep{prasad,prasad1}. The phase transition (PT) can have all sorts of observable signatures starting from gravitational wave generation to neutrino emission. The tilt angle or the angle between the rotation and the body symmetry axis will also change during the PT. Therefore, multimessenger astronomy comes into the forefront during the PT of nuclear matter (NM) to QM at NS's cores.
	
	PT in NS has been a subject of scrutiny for a long time. There has been much literature discussing the process and nature of PT \citep{abhijit,drago,mishustin,prasad,prasad1,irfan,amit,olinto,abhijit,jaikumar,alford15,drago15,ritam2020}. One of the models for this PT is a two-step PT where NM is first deconfining to up and down QM (also known as 2-flavor (2f) QM) \citep{abhijit,abhijit2,prasad,prasad1,ritam2020}. The next step is converting excess of down to strange quark for the QM to attain absolute stability (3-flavor matter (3f)). The weakly interacting conversion of down to strange quark generates a massive amount of neutrino and antineutrino pairs which being almost noninteracting, come out of the star and account for their neutrino emission \citep{ritam-neutrino,kovacs,abhijit-plb,sanjay,anand}.
	
	The process of PT is challenging to observe directly as it happens at the core of the NS, which is hidden from direct observation.
	However, there can be indirect observational signatures coming from them like the short bursts of gravitational wave signals \citep{prasad,prasad1,ritam2020} of a similar time scale of the PT and accompanied by neutrino-antineutrino signal. There can be an additional signature in the form of tilt angle evolution.
	As the NS converts to a quark star (QS) (sometimes also called a hybrid star due to the presence of both QM and NM in them), the star becomes more compact, and therefore the tilt axis of the star changes. As the PT is quite fast, the tilt angle also evolves quickly. Therefore, a sudden change of any observed pulsar's tilt angle can be thought to be the signature of PT happening at NS cores.
	
	In this article, we study the multimessenger observable signatures associated with nuclear to quark transition in NSs. In section 2, we describe our formalism for the multimessenger observation from the NS to QS conversion:  the neutrino energy deposition, the tilt angle evolution. 
	Results of the multimessenger observables are discussed in section 3, and finally, in section 4, we summarize our results and draw conclusions from them.
	
	\section{Formalism}
	\subsection{2f to 3f Conversion}
	Combustion in the NS due to shock is a two-step process: a sudden density fluctuation at NS cores which initiates a shock and deconfines NM to a 2f QM; as 2f QM is unstable gains absolute stability by converting to 3f QM. The first process is very fast and happens at the time scale of strong interaction \citep{prasad,prasad1}. Most of the observable signatures like neutrino emission and tilt angle evolution come from the second process (the 2f-3f conversion), and in this work, we mostly study these aspects. In the second process, the excess of down ($d$) quarks convert to strange ($s$) quarks as long as the chemical potential of down quarks exceeds strange quark's chemical potential and make 3f matter via following weak decays
	\begin{align}
	&d\rightarrow u+e^-+\bar{\nu}_e\\
	&u+e^-\rightarrow d+\nu_e\\
	&s\rightarrow u+e^-+\bar{\nu}_e\\
	&u+e^-\rightarrow s+\nu_e\\
	&d+u\leftrightarrow u+s
	\end{align}
	In the 3f beta equilibrated QM, the baryon conservation and charge neutrality conditions also holds, which is given by
	\begin{align}
	&2n_{u}=n_{d}+n_{s}+3n_{e^-} \\
	&3n_b=n_{d}+n_{s}+n_{u}
	\label{brn-cons}
	\end{align}
	where $n_b$, $n_{u}$, $n_{d}$, $n_{s}$ and $n_{e^-}$ are baryon, up quark, down quark, strange quark and electron number densities respectively. The combustion process  starts at the center of the star and goes towards the surface. The conversion of d to s quarks is governed by a parameter ``a'' defined as \citep{alford15,drago15,ritam2020}
	\begin{equation}\label{a}
	a(r)=\frac{n^{2f}_{k} (r)-n^{3f}_{k}}{n_b(r)}
	\end{equation}
where $n_k=\frac{1}{2} (n_d-n_s)$. Superscript 2f and 3f denotes the  2f and 3f matter respectively. $n^{3f}_k$ is asymptotically the value of $n_k$ in the 3f equilibrated matter which will be constant throughout the process and approximately close to $0$, the minimum value of $a$. At the conversion front we assume $n_k=n^*_k $, then  $a=\frac{n^*_k}{n^{3f}}=a^*$, where $`a^*$' can be found as $a^{*}=\sqrt{\dfrac{2\Delta\mu_b\chi^{Q}_k}{n^{3f}}}$,  where $\Delta\mu_b $ difference in the chemical potential across the front and the susceptibility $\chi^{Q}_k$ is defined as $\chi^{Q}_k =\frac{\partial n_k}{\partial (\mu_d-\mu_s)}$. For purely 2f matter, $n_k=n^{2f}_d/2=n^{2f}$ due to absence of $s$ quarks in this  region $a \approx \frac{n^{2f}}{n_{b}} \approx 1$ which is the maximum possible value of $a$. Therefore, $a$ lies in the range between $0$ to $1$.
	
	The decay of $d$ quarks into $s$ quarks at the conversion front and diffusion of $s$ quarks into the matter happens simultaneously. The differential equation (DE)  coupling these two processes is given by \citep{ritam2020}
	\begin{equation}
	D\frac{d^2 a}{d r^2}-\text{v}\frac{d a}{d r}-R(a)=0.\label{fnl- eq-x}
	\end{equation}
	where $D$ and $R(a)$ are the decay rate and the diffusion coefficient respectively and are given by \citep{olinto,abhijit}
	
	\begin{equation}
	D\simeq 10^{-3}\Bigg(\frac{\mu_b}{T}\Bigg)^{2}\: cm^2 s^{-1},\label{D}
	\end{equation}
	
	\begin{equation}\label{R}
	R(a)\simeq\frac{128}{27 \times 15\pi^3} G^2_F\cos^2\theta_c\sin^2\theta_c\mu^5\, a^3=\frac{a^3}{\tau},
	\end{equation}  
	with
	\begin{equation}
	\tau=\Big[\frac{128}{27 \times 15\pi^3} G^2_F\cos^2\theta_c\sin^2\theta_c\mu^5\Big]^{-1}\simeq 1.3\times 10^{-9}\Big[\frac{300\, MeV}{\mu}\Big]^5 s.
	\label{tau}
	\end{equation}
	Redefining $x$ as $x=\eta \xi$ and $\eta=\frac{D}{\text{v}}$, DE \ref{fnl- eq-x} is written as
	\begin{equation}\label{fnl-eq-xi}
	\frac{d^2 a}{d\xi^2}-\frac{d a}{d\xi}-g a^3=0.
	\end{equation}
	where 
	\begin{equation}\label{g}
	g\equiv\frac{D}{\tau \text{v}^2}.
	\end{equation}
	Integrating equation \ref{fnl-eq-xi} over small volume of cylinder whose axis coincides with $r$, and assuming $r\rightarrow 0$ we get boundary condition,
	\begin{equation}\label{bc2-1}
	\frac{d a}{d\xi}(0)\equiv\frac{d a}{d \xi}\Big|_0=-\Big(a^*-\frac{n^{2f}}{n_{b}}\Big).
	\end{equation}
	On one hand we know that a lies between $0$ and $1$ and on the other hand equation \ref{bc2-1} shows that $\frac{da}{d\xi}$ is always negative which means `$a$' is a monotonic decreasing function. Hence, it is clear that solution of DE \ref{fnl-eq-xi} will always be a monotonic decreasing function. Starting with a guess value of $g$ one can solve DE \ref{fnl-eq-xi}. If the guess value of $g$ is greater than actual value then the solution overshoots and if less then it undershoots. After finding the correct value of $g$ (say $g_s$, the solution of DE \ref{fnl-eq-xi}), the front velocity is given by 
	\begin{equation}\label{v}
	\text{v}=\sqrt{\frac{D}{\tau g_{s}}}.
	\end{equation}
	
	\subsection{Neutrino Energy Deposition}
	As the combustion front for the 2f-3f PT travels from the center to the surface of the star, at each time interval $dt$, the combustion front moves a distance $dr$ such that $\frac{dr}{dt}=\text{v}=\sqrt{\frac{D}{\tau g_{s}}}$. At any instance,  ($t=t_{ins}$), we have the 2f matter converting to 3f matter. Thus at each $t_{ins}$, we have a shell of radius $r=r_{ins}$ from which $\nu_e$ and $\bar{\nu_e}$ are being formed and gets annihilated releasing energy in the process. To find the energy deposited by the annihilation process, we define the "neutrinosphere" $R_{n}$ which is the mean free path of the $\nu_e-\bar{\nu_e}$  collision.\\\\
	We take $r_{ins}$ to be the instantaneous neutrinosphere which evolves with time through the interior of the star. We then define the energy deposition rate at some distance $r$ due to the neutrino sphere $R_n=r_{ins}$,  ($r>R_n$) as \citep{kovacs}
	\begin{align}
	\frac{dE}{dt}=\int\int f_{\nu}(p_\nu,r)f_{\bar{\nu}}(p_{\bar{\nu}},r)[\Lambda(\sigma,\text{v}_{\nu},\text{v}_{\bar{\nu}},\varepsilon)]d^3p_\nu d^3p_{\bar{\nu}}
	\label{eq1} 
	\end{align}
	where $f$ is the number density of the neutrinos in momentum space, $p$ is the momentum of the neutrinos and $\Lambda$ is a function which depends on the cross-section of the $\nu_e-\bar{\nu_e}$ collision $\sigma$, the velocity of the neutrinos $\text{v}_{\nu}$, $\text{v}_{\bar{\nu}}$ and the neutrino energies $\varepsilon$. The subscript $\nu$ and $\bar{\nu}$ stands for neutrino and anti-neutrino respectively.\\\\
	Each of the $\nu_e$ and $\bar{\nu_e}$ has a solid angle of emission $\Theta$. The momentum thus can be written as $p_\nu=\varepsilon_\nu\Theta$ and the volume element as $d^3p_\nu=\varepsilon_\nu^2d\varepsilon_\nu d\Theta$. The momentum integral thus decomposes into an energy integral and angular integral. The energy with which each $\nu_e$ and $\bar{\nu_e}$ is emitted depends on the temperature of the neutrinosphere. Assuming that the neutrinos are emitted isotropically throughout the neutrinosphere, the integral of equation \ref{eq1} becomes \citep{bethe}
	\begin{align}
	\frac{dE}{dt}=\kappa A(T)B(r,\theta)
	\label{eq1.5}
	\end{align}
	where $A(T)$ is a function of the temperature $T$ of the neutrinosphere and $B(r,\theta)$ is the function of the path taken by the $\nu_e$ and $\bar{\nu_e}$ to escape from the neutrinosphere and $\kappa$ is constant of proportionality. In order to calculate the function A and B, we start with the metric of a rotating star given by \citep{hartle}
	\begin{align}
	ds^2=-e^{2\nu(r,\theta)}dt^2+e^{2\lambda(r,\theta)}dr^2+e^{2\mu(r,\theta)}d\theta^2+e^{2\psi(r,\theta)}\Big[d\phi-\omega(r)dt\Big]^2
	\end{align}
	where $\nu$, $\lambda$, $\psi$ and $\mu$ are unknown functions of $r$ and $\theta$. For slow rotation, these coefficients can be written as
	\begin{align}
	&e^{2\nu(r,\theta)}=e^{2\Phi(r)}\Big[1+2\Big(h_0(r)+h_2(r)P_2(\cos\theta)\Big)\Big]\label{eq2}\\
	&e^{2\lambda(r,\theta)}=e^{2\Lambda(r)}\Big[1+\frac{2e^{2\Lambda}}{r}\Big(m_0(r,\theta)+m_2(r,\theta)P_2(\cos\theta)\Big)\Big]\label{eq3}\\
	&e^{2\mu(r,\theta)}=r^2\Big[1+2k_2(r)P_2(\cos\theta)\Big]\label{eq4}\\
	&e^{2\psi(r,\theta)}=r^2\sin^2\theta\Big[1+2k_2(r)P_2(\cos\theta)\Big]\label{eq5}
	\end{align}
	where $\Phi$ and $\Lambda$ are solved using TOV \citep{tov1,tov2} equations and $h_0$, $h_2$, $m_0$, $m_2$ and $k_2$ are functions of $r$ which can be solved numerically from center to the surface of the star using the hartle-throne DEs \citep{hartle}. Using the metric, the deflection angle of emission is calculated from the null geodesics as \citep{kovacs}
	\begin{align}
	&\Delta\phi\Big|_{\theta=\pi/2}= \nonumber\\
	&-\int_{r_{em}}^{r_{obs}}\frac{\sqrt{e^{2 \lambda (r)}} \left(b e^{2 \nu (r)}+e^{2 \psi (r)} \omega (r)\right)}{\sqrt{\left(e^{2 (\nu (r)+\psi (r))}+e^{4 \psi (r)} \omega (r)^2\right) \left(e^{2 \psi (r)} (1-2 b \omega (r))-b^2 e^{2 \nu (r)}\right)}} \label{eq55}
	\end{align}
	Where $b$ is the impact parameter. This deflection angle is then used to calculate the solid angle $d\Theta$ integral for the emission of neutrino. Thus at the equatorial region the expression for $B(r,\pi/2)$ using equation \ref{eq2} - \ref{eq55} gives
	\begin{align}
	B(r,\pi/2)=\frac{2\pi^2}{3}\Big[6+\Upsilon(r)+4\sqrt{1+\Upsilon(r)}\Big]\Big[\sqrt{1+\Upsilon(r)}-1\Big]^4
	\label{eq6}
	\end{align}
	where,
	\begin{align}
	&\Upsilon(r)\Big|_{\theta=\pi/2}=\nonumber\\
	&\frac{e^{4 \psi (R_n)-2 \psi (r)} \left(-e^{2 \nu (r)}-e^{2 \psi (r)} \omega (r)^2\right)}{\left(e^{2 \psi (R_n)} (\omega (r)-\omega (R_n))+\sqrt{e^{2 (\nu (R_n)+\psi (R_n))}+e^{4 \psi (R_n)} \omega (R_n)^2}\right)^2}
	\end{align}
	Thus $B$ defines the path of the neutrino emission. Now in order to calculate $A$, we at first define the gravitational redshift parameter by comparing the proper time $d\tau$ with the coordinate time $dt$ as
	\begin{align}
	d\tau^2=\Big[e^{2\nu(r)}+\omega(r)^2e^{2\psi(r)}\Big]dt^2
	\end{align}
	The number density can be written as
	\begin{align}
	f(p_\nu,r)\propto\frac{1}{1+e^{\varepsilon_\nu/k_BT(r)}}.
	\end{align}
	Thus the function $A(T)$ is evaluated by the integrating $f_{\nu}(p_\nu,r)f_{\bar{\nu}}(p_{\bar{\nu}},r)\varepsilon_\nu^2\varepsilon_{\bar{\nu}}^2[\Lambda(\sigma,v,\varepsilon)]d\varepsilon_\nu d\varepsilon_{\bar{\nu}}$ over the energies and we get \citep{asano,kovacs}
	\begin{align}
	A(T)\propto (k_BT(r))^9
	\label{eq7}
	\end{align}
	The $T$ at neutrinosphere $R_n$ is related to $T$ at a distance $r$ by the equation
	\citep{kovacs}
	\begin{align}
	\frac{T(r)}{T(R_n)}=\frac{\lambda(R_n)}{\lambda(r)}=\frac{\Big[e^{2\nu(R_n)}+\omega(R_n)^2e^{2\psi(R_n)}\Big]^{1/2}\lambda(R_n\rightarrow\infty)}{\Big[e^{2\nu(r)}+\omega(r)^2e^{2\psi(r)}\Big]^{1/2}\lambda(r\rightarrow\infty)}
	\label{eq8}
	\end{align}
	The luminosity at infinity which is observable can be related to the luminosity at the neutrino sphere as \citep{kovacs},
	\begin{align}
	&L(r\rightarrow{\infty})=\Big[e^{2\nu(R_n)}+\omega(R_n)^2e^{2\psi(R_n)}\Big]L(R_n)\nonumber\\
	&=\Big[e^{2\nu(R_n)}+\omega(R_n)^2e^{2\psi(R_n)}\Big]4\pi R_n^2\sigma_B T(R_n)^4
	\label{eq9}
	\end{align}
	where $\sigma_B$ is the stefan-Boltzmann constant. Combining equation \ref{eq1.5}, \ref{eq6}, \ref{eq7}, \ref{eq8}, and inverting equation \ref{eq9} to solve for $T$ in terms of luminosity at infinity the energy deposition rate for neutrinos is given by 
	\begin{align}
	&\frac{dE}{dt}=\kappa\Bigg[\left(\frac{e^{2 \nu (R_n)}+e^{2 \psi (R_n)} \omega (R_n)^2}{e^{2 \nu (r)}+e^{2 \psi (r)} \omega (r)^2}\right)^{9/2}\nonumber\\
	&\left(\frac{L(r\rightarrow\infty)}{\pi  R_n^2 \sigma  e^{2 \nu (R_n)}+\pi  R_n^2 \sigma  e^{2 \psi (R_n)} \omega (R_n)^2}\right)^{9/4}\Bigg]\nonumber\\
	&\Big[6+\Upsilon(r)+4\sqrt{1+\Upsilon(r)}\Big]\Big[\sqrt{1+\Upsilon(r)}-1\Big]^4
	\label{dep}.
	\end{align}

	\subsection{Tilt Evolution}
	During the PT, along with the neutrino emission, the star loses energy, and its structure also changes. Therefore, the misalignment angle also changes.
	If the star has some axisymmetry along some particular axis due to the magnetic field, then we can write the moment of inertia for such a system as \citep{lander1}
	\begin{align}
	I_{ij}=I_0\delta_{ij}+\Delta I\left(n_in_j-\frac{\delta_{ij}}{3}\right)
	\end{align}
	where the unit vector $n_i=(0,0,1)$ points in the direction of the deformation axis. The two principal moments $I_1=I_2$ are equal and the third moment defines the deformation $\Delta I=I_3-I_1$. Thus $I_0$ is the moment of inertia for spherical symmetric case when $I_1=I_2=I_3$ and $\Delta I=0$.\\
	Thus in general if the angular velocity of the star is $\Omega_i=(\Omega_1,\Omega_2,\Omega_3)$ then we can calculate the angular momentum as
	\begin{align}
	J_i=I_{ij}\Omega_j=\left(I_0-\frac{1}{3}\Delta I\right)\Omega_i+\Delta I \Omega_3n_i
	\label{angular}
	\end{align}
	The angle $\chi$ between $J_i$ and $n_i$ is the misalignment angle of the star. The total amount of energy that has to be taken from the star to change the misaligned angle from $\chi_1$ to $\chi_2$ is
	\begin{align}
	E=\int_{\chi_1}^{\chi_2}\frac{dJ}{dt}d\chi
	\label{work}
	\end{align}
	Equation \ref{work} can be inverted and using equation \ref{angular} we can write \citep{lander2}
	\begin{align}
	\dot{\chi}=\frac{\dot{E}}{I\Omega^2}\cot\chi
	\label{mis}
	\end{align}
	Thus as the star loses energy, the misalignment angle evolves.
	
	\begin{table*}
		\centering
		\caption{PT in 2f star}
		\begin{footnotesize}
			\begin{tabular}{@{\extracolsep{2pt}}ccccccccccc}
				\hline 
				\hline
				\multicolumn{3}{c}{2f star}&  \multicolumn{4}{c}{3f star} & PT time\\ 
				\cline{1-4} \cline{5-8} \cline{9-10}
				$\rho_{c}$& $M_{G}$ &$ R_{e}$ & $\rho_{c}$ &$\nu $&$M_{G}$ & $R_{e}$ & t \\
				$(10^{14} g/cc)$ &  $(M_{\odot})$& (km) & $(10^{14} g/cc)$ & (Hz) &  $(M_{\odot})$& (km) & (ms)  \\
				\hline
				6.65 & 1.6043 & 13.74 & 7.77 & 51.7 & 1.6010 & 13.26 &  2.23   \\
				7.70& 1.8008 & 13.67 &9.40 & 52.8 & 1.7955 & 13.08 &   2.44  \\
				9.70& 2.0013 & 13.44  &13.95& 56.3 & 1.9927 & 12.55 &  4.47 &  \\
				\hline
			\end{tabular} 
		\end{footnotesize}
	\end{table*}

	
	\section{Results and Discussions}
\begin{figure}[h!]
		\centering
		\includegraphics[scale=0.37]{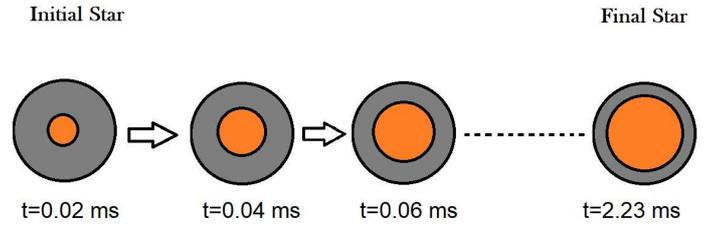}
		\caption{Evolution of the combustion front as 2f matter are converted to 3f matter. The time the process takes is shown for the $1.6$ $M_{\odot}$ star. Each of the intermidiate star (star 'n', where n=1,2,3.....120) has a fixed radius upto which the 2f matter has been converted to 3f matter. Also each star n is associated with a time $t_{ins}$ which is the instantaneous time of the ongoing dynamic PT.}
		\label{PT}
	\end{figure}

	We start our calculation with S271 \citep{lala,horo} parameter setting to describe the NM and MIT bag model having quark interaction \citep{chodos,alford2,weissenborn} to describe QM. 
	These EoS are in agreement with the recent nuclear and astrophysical bounds. The 2f matter consists only of up and down quarks (with masses $2$ and $5$ MeV respectively),  whereas the 3f QM additionally has strange quarks in them of mass $95$ MeV. The bag constant is taken to be $B^{1/4}=140$ MeV, and the quark coupling value $a_4$ is $0.5$.
	The deconfinement of NM to 2f QM is almost instantaneous; however, the 2f to 3f conversion is relatively slow. Most of the observable signature comes from the second process.
	We construct a 2f star with a given central density $\rho_{c}^{2f}$ having a baryonic mass $M_{B}$ with the rotational frequency of 50 Hz; we then go on to construct a 3f star with a central density $\rho_{c}^{3f}$ such that it has the same baryonic mass as 2f star. The 3f star has a higher central density, a smaller radius, and slightly larger rotational velocity than the 2f star. We carry out our studies on a cold 2f star and assume the $T$ to be $10^{-2} MeV$. The $T$ does not affect the EoS significantly but governs the diffusion dynamics of the 2f to 3f quark conversion (equation \ref{D}). We perform our calculation as defined in section 2.1 and obtain the PT front velocity. Using the conversion velocity, we find the change in the star's density profile as a function of time as 2f matter settles into 3f QM. We then construct stable intermediate stars with the baryonic mass of the star kept fixed while ensuring conditions that the 3f matter is present from the center up-to-the-front location, followed by purely 2f matter up to the critical point (few times of nuclear saturation density), and finally, NM is present in the outer region. We have constructed approximately 120 such intermediate star (for $1.6$ $M_{\odot}$ star). Each of these intermediate stars represents the NS state at a particular instant of time, while it undergoes PT from 2f to 3f matter. The intermediate star is constructed in such a way that at a particular instant of time (that is, at a particular intermediate star $n$, where $n=1,2,3,4......120$), the radius of combustion (neutrinosphere) has increased up to some $r=r+dr=R_n$ (the conversion velocity) as shown in figure \ref{PT}. \\

	The conversion velocity increases from the center to the star's surface, and its magnitude is of the order of $ \sim 10^{-2}$ times the speed of light. It takes about a few milliseconds for the PT process to complete. The PT timescale is seen to be directly proportional to the mass of the initial 2f star. The results are presented in Table 1 for three different masses of 2f stars. The PT process described here is associated with the emission of neutrinos, and the newly formed 3f star has different rotational properties, and these aspects are studied below.

\begin{figure}[h!]
		\includegraphics[scale=0.55]{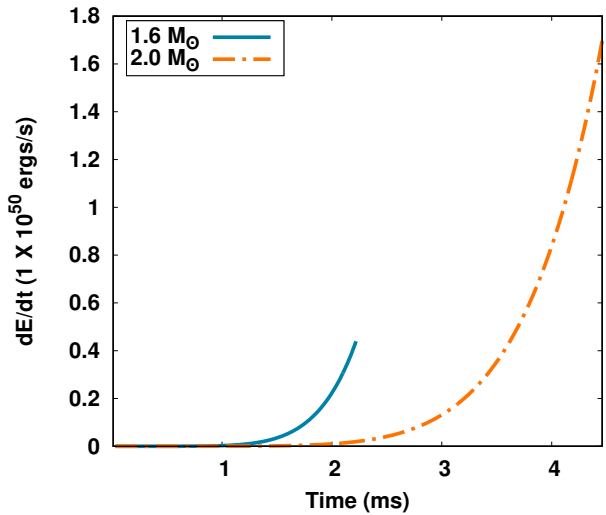}
		\caption{Energy deposition rate on the surface of the QS has been plotted as a function time. The y-axis of the star shows the total amount of energy that has been deposited on the surface of the star at that instant of time. The total energy has been normalized by $10^{50}$ $ergs$. At $t=2.23$, $4.47$, value of $dE/dt$ gives the total energy that has been emitted throughout the process for $1.6$ $M_{\odot}$, $2.0$ $M_{\odot}$ stars respectively. }
		\label{edep}
	\end{figure}

	The total energy deposition rate is calculated by numerically solving equation \ref{dep}. The energy deposition equation has been numerically integrated from $t(r_i)=0$ to $t(r_f)=t_f$ dynamically by using the profile of each intermediate star for each time instant. Here $r_i$ and $r_f$ are the initial and final radius of the combustion front (neutrinosphere) and $t_f$ is the final time for PT. The energy deposition rate is calculated and plotted as a function of time for two masses in figure \ref{edep}. The y-axis of the plot shows the amount of energy deposited on the star's surface by the neutrinos generated by weak combustion at the neutrinosphere at that particular time. The final energy that is deposited is shown in Table \ref{tab1}.
	It is seen that the energy deposited is in the range $10^{49}$ ergs to $10^{50}$ ergs, and the time scale for the process lies between $2.23$ ms to $4.47$ ms. 
	
	\begin{table}
		\caption{Table showing the total energy deposition of neutrino-antineutrino annihilation on the surface of the star.}
		\begin{tabular}{ |p{1.7cm}|p{2.5cm}|p{2.5cm}| }
			\hline
			\multicolumn{3}{|c|}{Energy Deposition Rate} \\
			\hline
			Mass ($M_{\odot}$)&Total Energy (ergs)&Total Time (ms)\\
			\hline
			\hline
			\vfill$1.6$ & \vfill$4.4\times10^{49}$&\vfill 2.23\\
			$1.8$ & $7.6\times10^{49}$ & 2.43\\
			$2.0$ & $1.7\times10^{50}$ & 4.47\\
			\hline
			
		\end{tabular}
		\label{tab1}
	\end{table}
	
	Astrophysically, many events of Short Gamma-Ray Burst (SGRB) have been detected \citep{nakar}. One such event is the GRB 050709 \citep{fox}. At a redshift of $z=0.16$, the isotropic equivalent radiated energy in the rest frame of the source was calculated to be $E_{iso}\approx6.9\times 10^{49}$ $ergs$. The work of \citep{villasenor} has eliminated many possible sources of such SGRBs. The source was observed to be at an offset with the galaxy's center; hence the possibility of Soft Gamma-ray Repeaters (SGR) giant flares was eliminated. Another possibility was the gravitational collapse of a massive star that could lead to supernova and produce similar energy signatures; however, the collapse models predicted time signatures more than few seconds, and hence such sources were also eliminated.
	
\begin{figure}[h!]
		\centering
		\includegraphics[scale=0.21]{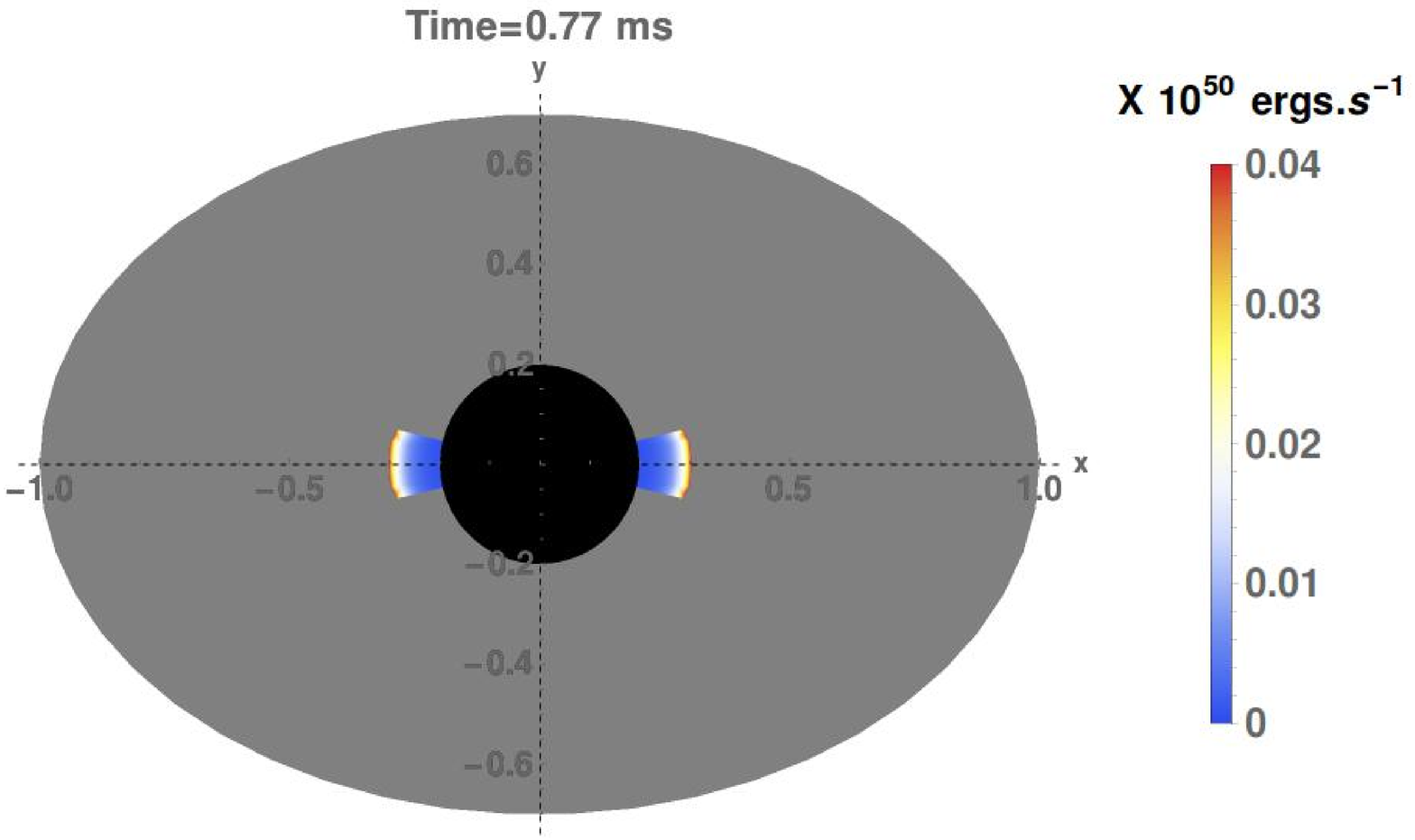}
		\includegraphics[scale=0.21]{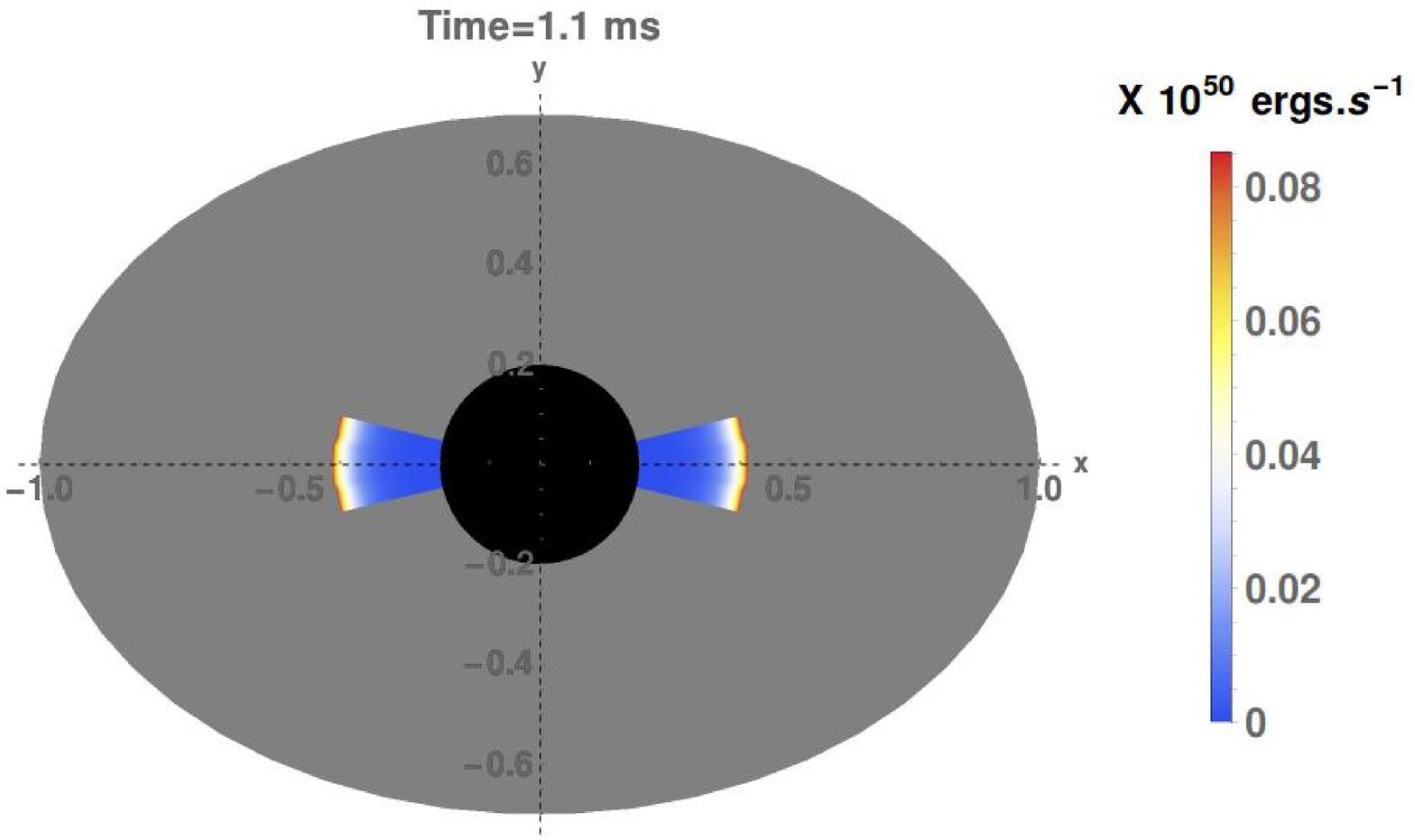}
		\includegraphics[scale=0.21]{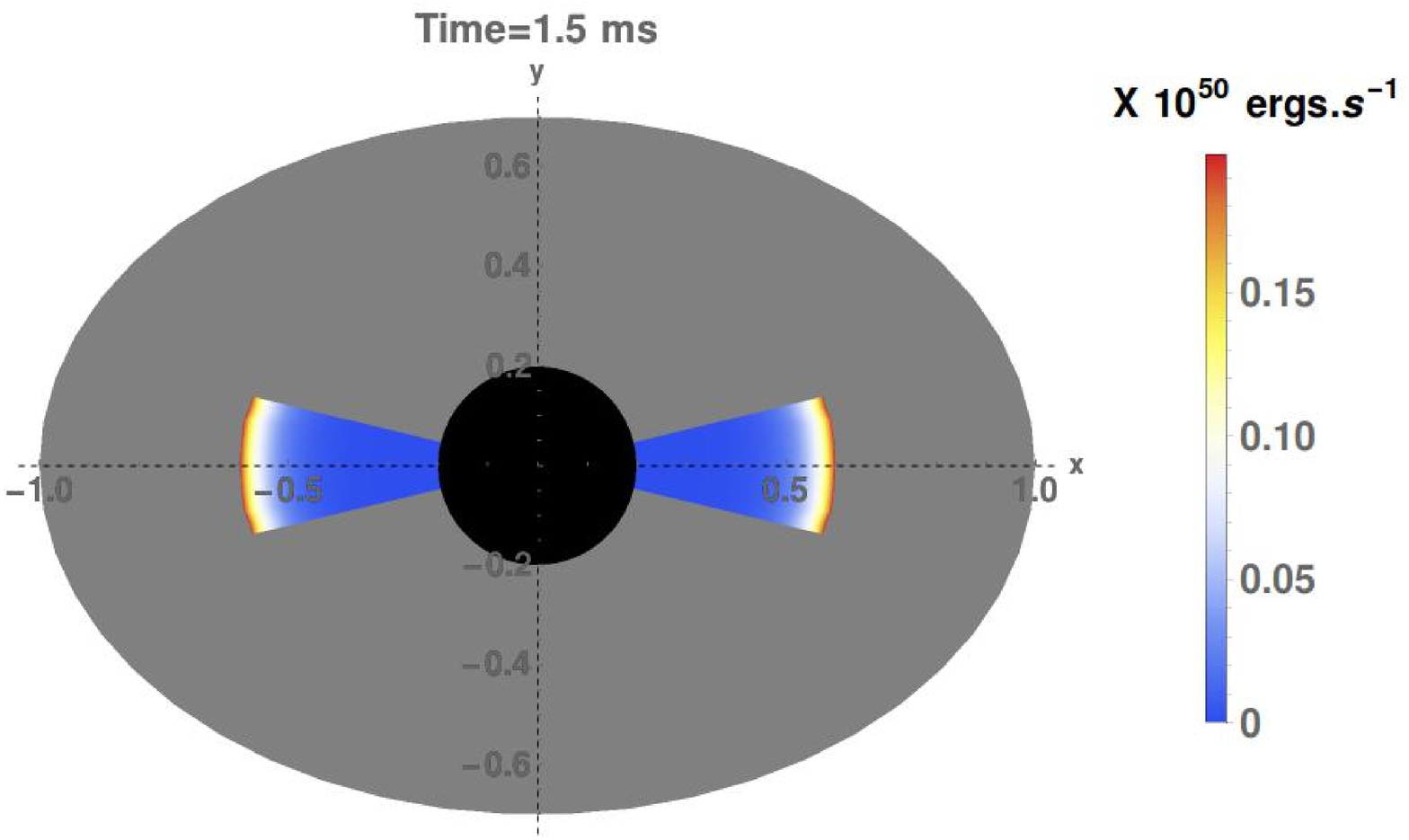}
		\includegraphics[scale=0.21]{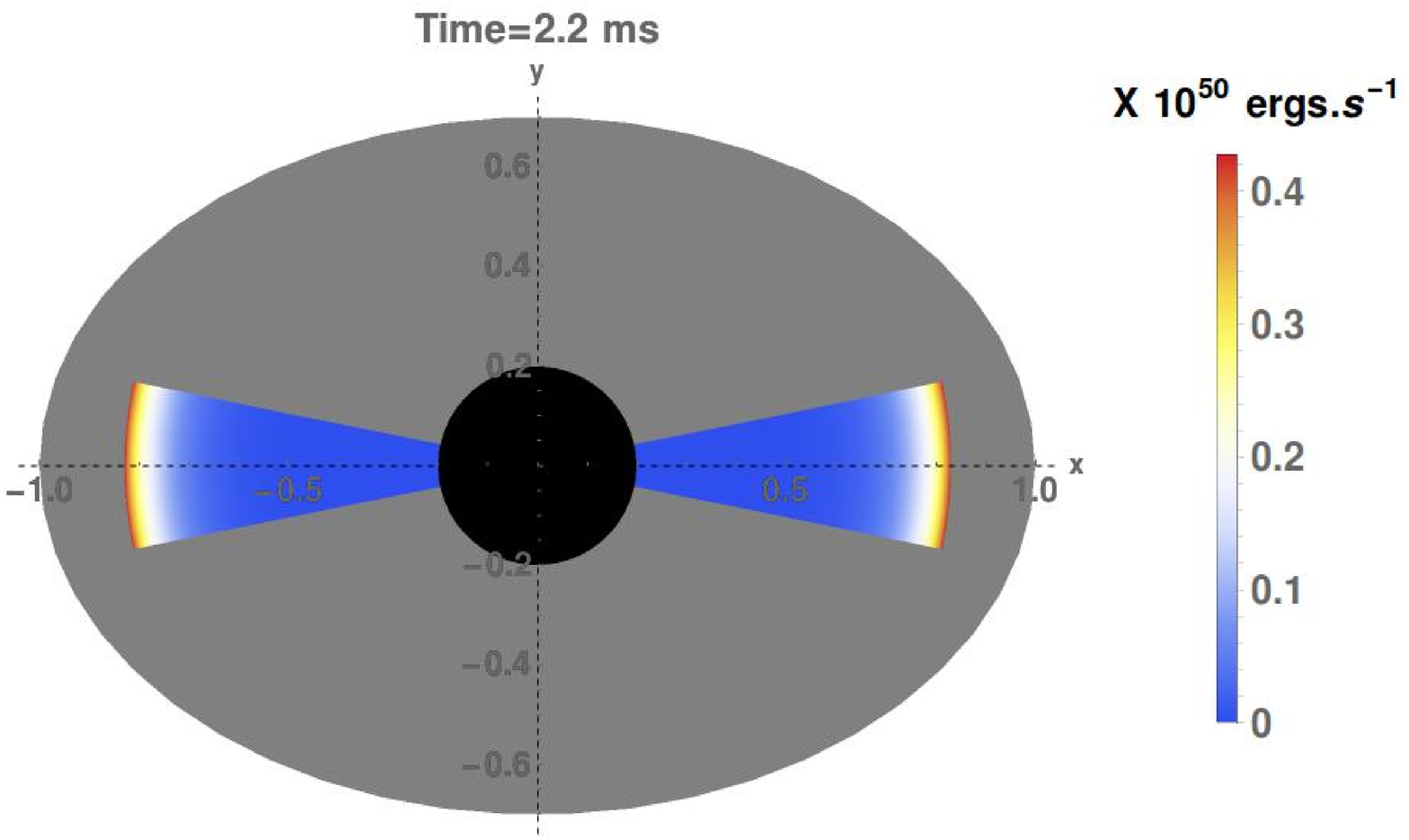}
		\caption{Snapshot of simulation of dynamic evolution of neutrino and antineutrino escape and annihilation has been shown. At each snapshot, the neutrino and antineutrino can be seen produced due to PT and gets annihilated. The corresponding energy deposited on the surface at each radius is shown as a heat map. The simulation only shows deposition along $\theta=\pi/2$ (equatorial region). The dark sphere is the opaque region for neutrinos.}
		\label{escape}
	\end{figure}

	A possible source of such SGRB has been credited to NS-NS mergers, where the energy and time signatures from the models fall within the observed limit. However, this does not seclude the possibility of observations from isolated stars undergoing PT, as shown in our model. The energy and time signatures of neutrino-antineutrino annihilation from the PT process fall well within the observed limit. We have calculated this neutrino escape and annihilation process in the simulation snapshot shown in figure \ref{escape}. The simulation shows only the emission through the equatorial region because the geodesic calculated in equation \ref{eq6} holds only for the region $\theta=\pi/2$. Also, we have neglected the contribution of energy deposition due to a core of $\approx$ $2$ km where densities are much higher, and thus the QS is opaque to neutrinos in this region \citep{alcain}. \\
	
	However, the neutrino should be emitted and annihilated from all values of $\theta$. Nevertheless, it has been argued \citep{kovacs} that the energy emitted in the equatorial region gives a good approximation in terms of the order of the total energy emitted. This is because the integral in equation \ref{eq1} is decomposed into angular integral and energy integral.  The angular integration is restricted for $\theta=\pi/2$ due to the axisymmetry of the system. The angular integral thus only contributes to the path taken for neutrino emission; however, the neutrinosphere's energy integral contributes to the order of energy deposition due to neutrino-antineutrino annihilation. It could be seen in Table \ref{tab1} that, as the mass of the star increases, the time taken to convert 2f to 3f matter increases. Thus more and more neutrino will be emitted for higher mass stars, and thus the probability of collision of neutrino-antineutrino pair increases, and thus energy deposited increases. Hence models with a mass lower than $1.6$ $M_{\odot}$ and higher than $2.0$ $M_{\odot}$ will emit lower energy (with lower total time) and higher energy (with higher total time), respectively. Thus overall, with the model of dynamic conversion of NM-QM matter, SGRB events such as GRB 050509B (at $z\approx0.225$, $E_{iso}\approx4.5\times10^{48}$ $ergs$ \citep{gehrels,bloom}, GRB 060502B (at $z\approx0.287$, $E_{iso}\approx8.0\times10^{48}$ $ergs$ \citep{bloom2} could be explained. Detection of SGRB having pulses with $t_{90} \sim$ $ms$ ($t_{90}$ is the time duration in which $90\%$ of the photons detected are emitted from the source \citep{scargle}) can be explained particularly well by the PT model.\\
	
\begin{figure}[h!]
		\includegraphics[scale=0.55]{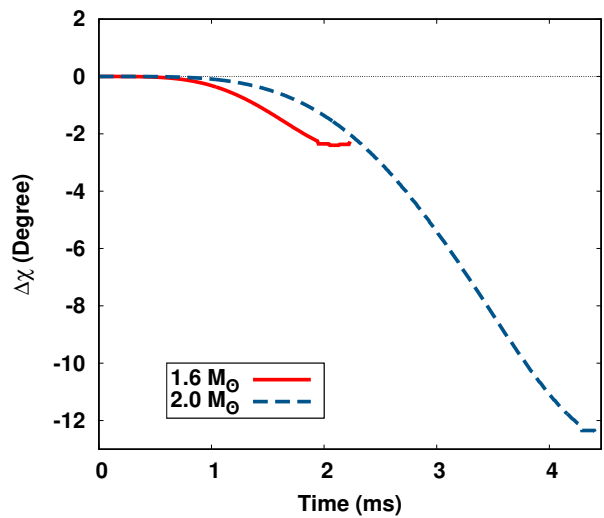}
		\caption{Misalignment angle evolution due to energy emitted from the star for three different masses. The misaligned angle is seen to be decreasing and going towards alignment. The more massive star has a higher probability of attaining alignment in comparison to the lower mass star. Here the initial angle is taken to be $\pi/5$ $(36^o)$.}
		\label{evo}
	\end{figure}

	As the NS loses a considerable amount of energy due to the 2f-3f conversion over a period of time, we have calculated how the NS's misalignment angle will evolve using equation \ref{mis}. The evolution has been plotted as a function of time as shown in figure \ref{evo} for two different masses ($1.6$ $M_{\odot}$ and $2.0$ $M_{\odot}$) from which energy is being emitted. \\
	
\begin{figure}[!tbp]
		\begin{minipage}[b]{0.2\textwidth}
			\includegraphics[scale=0.35]{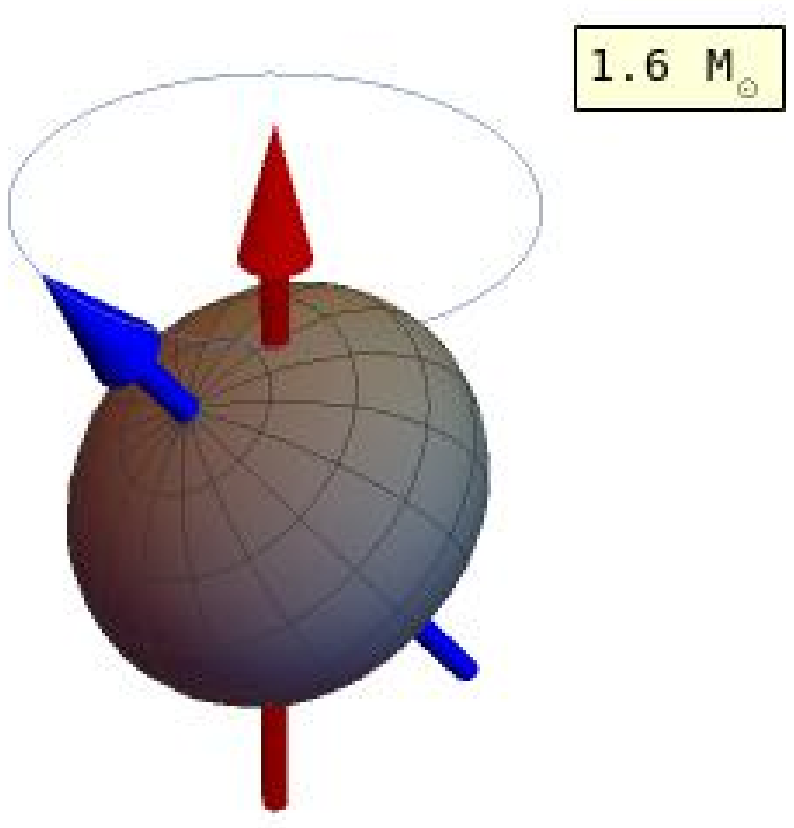}
			\includegraphics[scale=0.35]{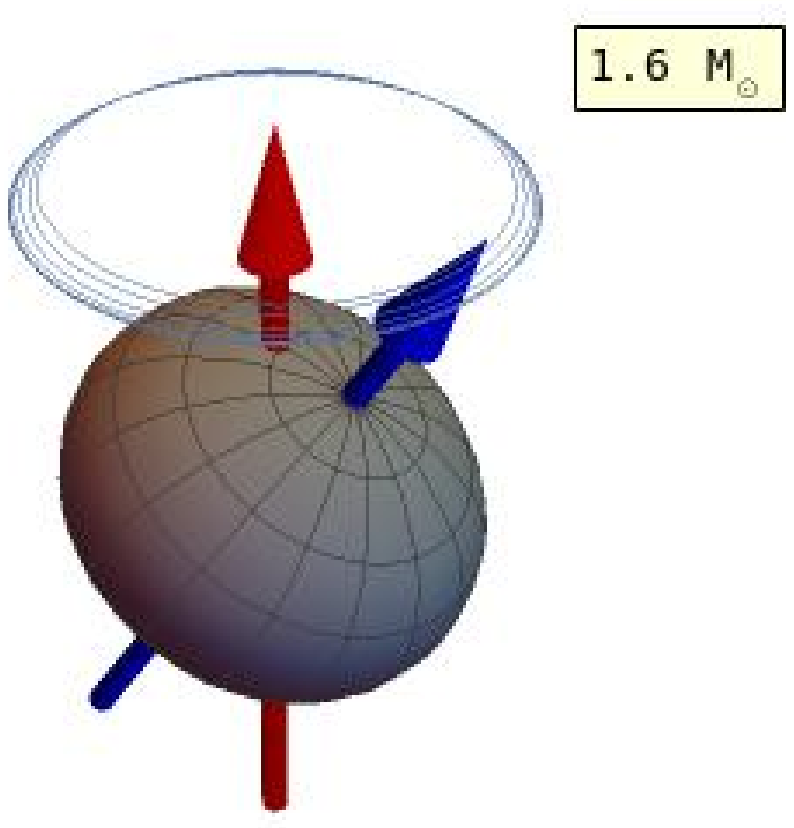}
		\end{minipage}
		\hfill
		\begin{minipage}[b]{0.25\textwidth}
			\includegraphics[scale=0.35]{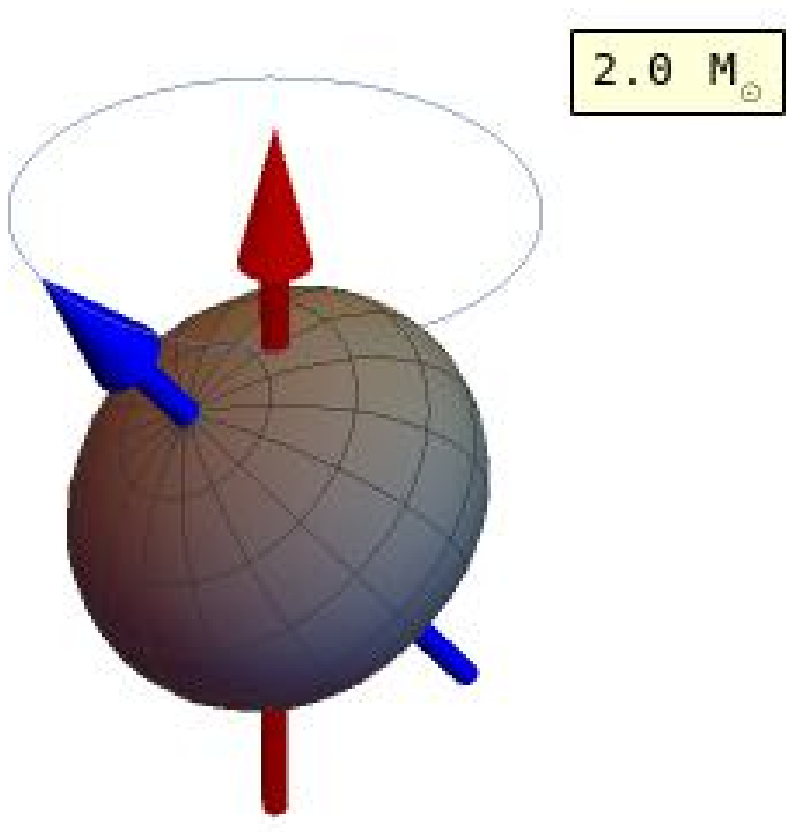}
			\includegraphics[scale=0.35]{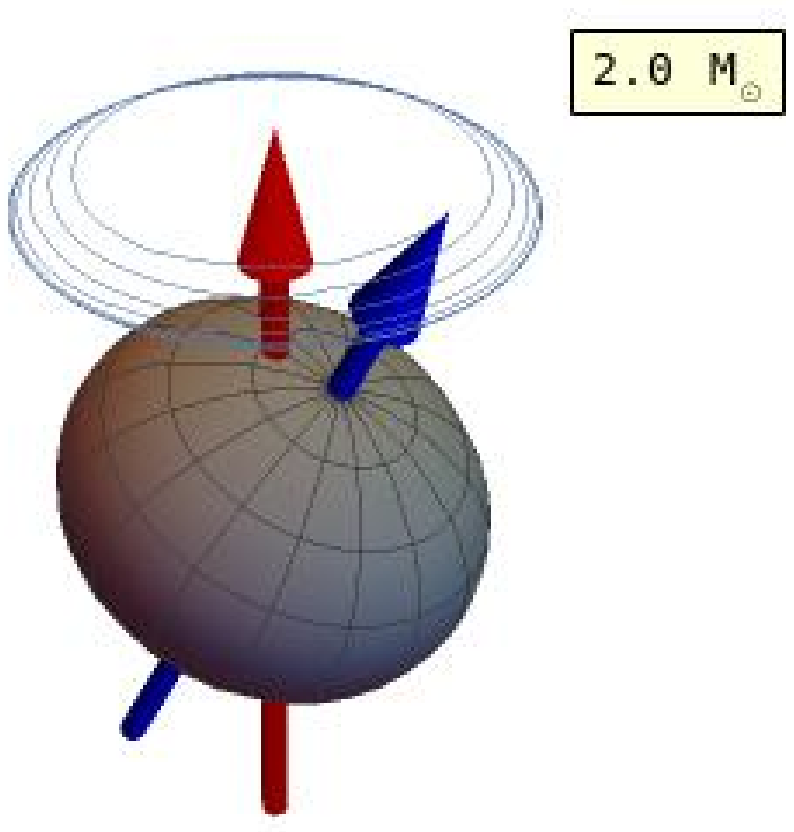}
		\end{minipage}
		\caption{Snapshot of evolution of angle can be seen for the beginning and end state of the star with two different masses. This simulation has been started with an initial angle $\chi_0=\pi/5$ $(36^{o})$. The final state of the star could be seen reaching alignment as the mass of the star increases left to right. The two star reaches its final angle after different times. ($10\pi/53\;(34^o)$ and $10\pi/75\;(24^o)$ for $1.6$ $M_{\odot}$ and $2.0$ $M_{\odot}$ respectively.)}
		\label{miss}
	\end{figure}

	The y-axis of the plot shows how much the misalignment angle $\chi$ changes with respect to some initial angle $\chi_0$. Thus the change $\Delta \chi$ is defined as $\Delta\chi\equiv\chi-\chi_0$. It could be seen from the figure \ref{evo} that the value of $\Delta\chi$ takes negative values; that is, the misalignment angle decreases as energy is being emitted from the star. It could be seen that the star could reach alignment if the initial angle $\chi_0$ is smaller than $\pi/9$. The previous spin-down model predicts alignment ranging from few seconds to years \citep{lander2}. In contrast, in our model, the time scales are in milliseconds and are purely due to the PT. Thus PT can, in general, quicken the alignment process of a misaligned star. This could be seen from the figure \ref{miss}, which is a snapshot of a simulation that shows the star's initial and final state during PT. As the star's mass increases, the star loses more amount of energy, and the time taken for the process increases with mass, and hence the value of $\Delta\chi$ increases. Thus more massive stars have a higher chance of getting aligned. \\
	
\begin{figure}[h!]
		\centering
		\includegraphics[scale=0.6]{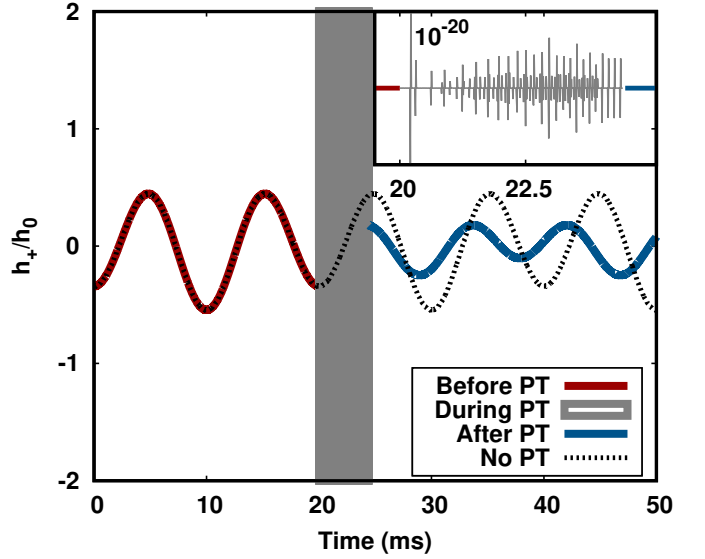}
		\caption{The GW signal template emitted from rotating NSs due to PT is shown in the figure. The y-axis shows the $h_+$ polarization of the gravitational wave normalized with the amplitude $h_0$. The distance from the earth's surface is taken to be $2$ $kpc$ and the line of sight angle is taken to be $30^o$. The first, second, and third gravitational wave plots are for a $2.0$ $M_{\odot}$ star where the amplitude is shown for regions, before, during, and after PT with no PT amplitude in the background for reference.}
		\label{gw}
	\end{figure}

	The PT process brings a change in the interior of the star. The density evolution during the settling of 2f star to a stable 3f star leads to quadrupole moment variation, leading to gravitational wave emission directly from the PT process itself, calculation of which is done in our latest work \citep{ritam2020}. The gravitational wave strain comes out to be of the order of $10^{-23}-10^{-21}$ for a source located at 1Mpc distance, and the amplitude spectrum calculation reveals peaks in the $1-10$ kHz range. We also provided the overall picture of possible gravitational wave emission. The initial neutron stars and final quark stars due to ellipticities can produce continuous GW emission, separated by short-lived signals originating from PT (for details, refer \citep{ritam2020}). The continuous gravitational wave signal coming from the star would also change due to the misalignment angle's evolution. To calculate the continuous waveform, we calculate the $h_+$ polarization \citep{bonazzola} given by
	
	\begin{align}
	h_+=h_0\sin\chi\left[\frac{1}{2}\cos\chi\sin{i}\cos{i}\cos\Omega t-\sin\chi\frac{1+\cos^2i}{2}\cos2\Omega t\right]
	\label{wave}
	\end{align}
	where,
	\begin{align}
	h_0=\frac{4G}{c^4}\frac{I\epsilon}{l}\Omega^2
	\end{align}
	where $i$ is the line of sight inclination and $\epsilon$ is the ellipticity of the star and $l$ is the radial distance from earth. We perform the GW calculation assuming that the source star at a distance $l=2$ $kpc$ (similar distance to crab pulsar) from the earth and at an inclination angle of $i=30^{o}$ and the ellipticity of NS and QS to be $10^{-4}$.
	The total waveform consists of continuous GW signals emitted from the NS and QS respectively (of the order of $10^{-27}$ and will change slightly due to the change in the misalignment angle and star structure due to PT) separated by a strong GW signal from the PT process itself (of the order of $10^{-19}$). 
	This is shown in figure \ref{gw}. 
	
	 It can be seen that the continuous gravitational wave before and after the PT changes both in amplitude and frequency. The amplitude decreases with a decrease of $\Delta\chi$ as it moves towards alignment, and the frequency increases because after the PT, the NS changes to QS. The QS becomes smaller than the NS due to the softening of the equation state, and hence the angular velocity of the QS increase. This results in an increase in the continuous gravitational-wave frequency due to PT but with a lower amplitude than the NS. It can be seen from equation \ref{wave}, the frequency of the gravitational wave is a combination of $\Omega$ and $2\Omega$; however, we see that as the star reaches alignment, only the $\Omega$ frequency will dominate.
	
	The gravitational wave amplitude emitted before the PT process for a star with initial $\Omega=314$ $rad$ $s^{-1}$ and $l=2$ $kpc$,
	\begin{align*}
	h_0=2.38\times10^{-27} 
	\end{align*}
	and after PT the final amplitude and frequency becomes, 
	\begin{align*}
	h_0=2.14\times10^{-27}\;\;\;\;\;\;\;\left( \Omega=325.27\,rad\,s^{-1}\right)\;\;\;\;\;\;\;(1.6\,M_{\odot})\\
	h_0=2.02\times10^{-27}\;\;\;\;\;\;\;\left( \Omega=331.70\,rad\,s^{-1}\right)\;\;\;\;\;\;\;(1.8\,M_{\odot})\\
	h_0=1.78\times10^{-27}\;\;\;\;\;\;\;\left( \Omega=351.84\,rad\,s^{-1}\right)\;\;\;\;\;\;\;(2.0\,M_{\odot}).
	\end{align*}
	
	However in general the $h_0$ for continuous gravitational waves can range between $10^{-27}-10^{-31}$ depending on the distance of the source form earth's surface $l$. To the present detectors, the continuous GW signals is still beyond detection capability however the strong signal from the PT process itself is well within the range of present detectors.
	
\begin{figure}[!tbp]
		\begin{minipage}[b]{0.12\textwidth}
			\centering
			\includegraphics[scale=0.50]{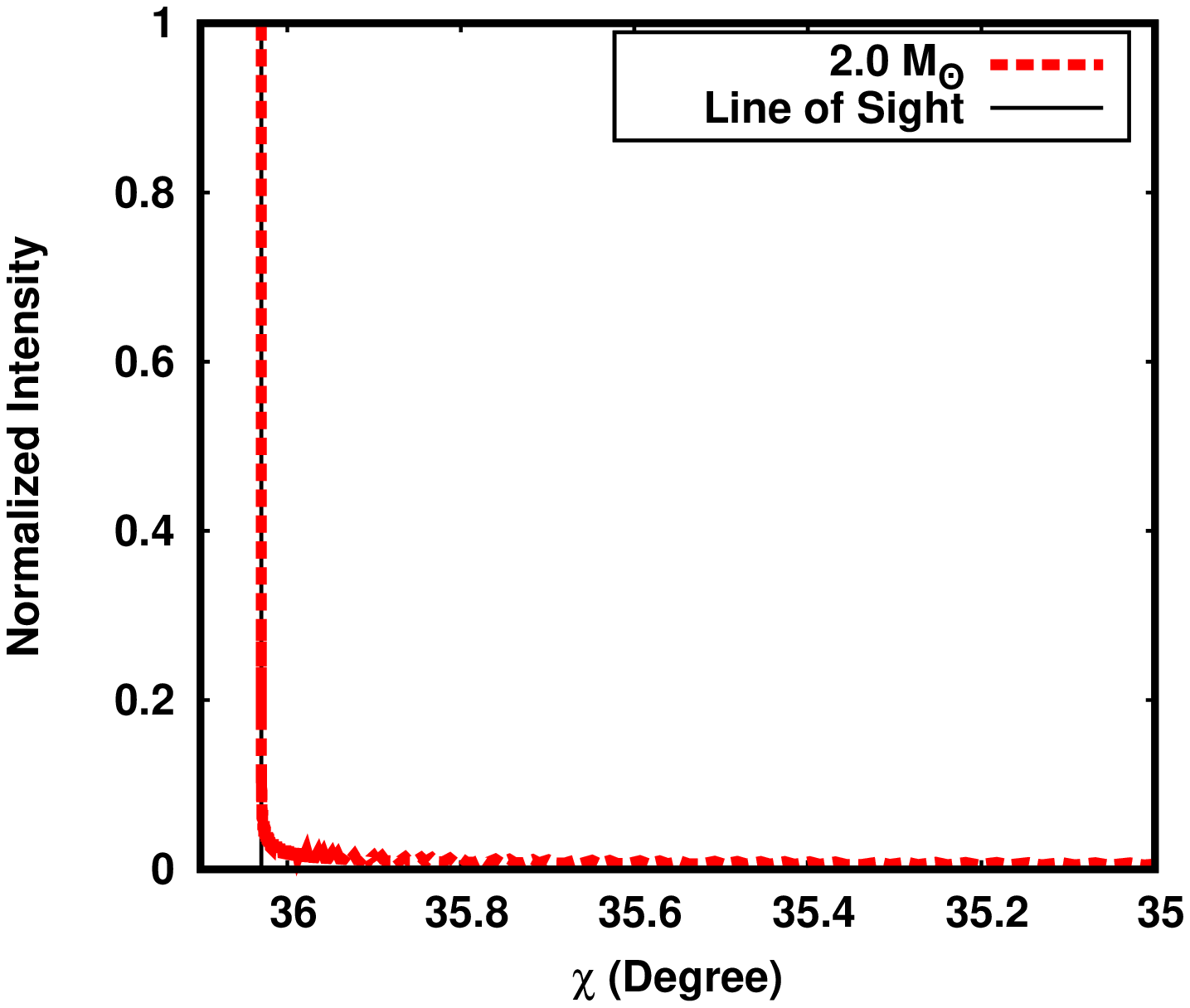}
			\subcaption{}
		\end{minipage}
		\hfill
		\hfill
		
		\begin{minipage}[b]{0.12\textwidth}
			\centering
			\includegraphics[scale=0.50]{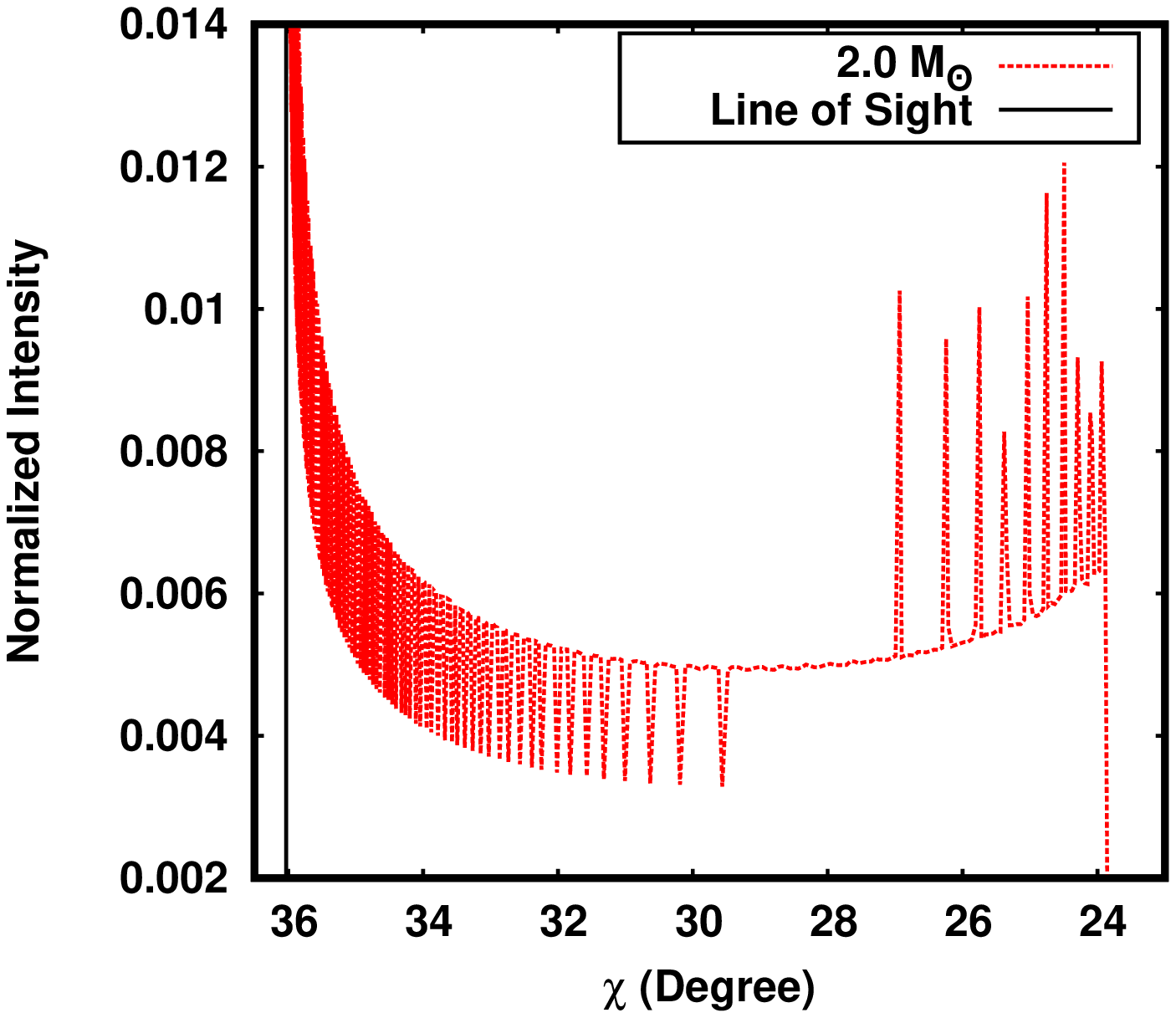}
			\subcaption{}
		\end{minipage}
		\caption{Normalized Intensity of radio emission from a $2.0$ $M_{\odot}$ radio pulsar has been plotted with respect to the evolution of misalignment angle of the pulsar. The $P$ for the pulsar evolves from $20.0$ $ms$ to $17.8$ $ms$ in $4.47$ $ms$ of PT time. (a) The intensity drops sharply as the misalignment angle passes through the line of sight (here from $\chi=\pi/5$ ($36^o$) to $\chi=10\pi/51$ ($35^o$)). (b) The intensity stabilizes to a value few order of magnitude lower than the peak intensity (for $\chi=10\pi/51$ ($36^o$) to $\chi=10\pi/75$ ($24^o$)). This can cause the pulsar signal to disappear or appear depending on the line of sight.}
		\label{int}
	\end{figure}

	The change in misalignment angle can also be reflected in electromagnetic emissions, such as the evolution of radio intensity from radio pulsars. Intensity from pulsars due to the presence of magnetosphere is given by \citep{wang}
	\begin{align}
	I_{norm}=P^{q-4}\dot{P}\cos^2{\chi}f^{q-3}\rho^{2q-6}
	\end{align}
	where $I_{norm}$ is the normalized intensity, $P$ and $\dot{P}$ are the period and the period derivative of the pulsar, the parameter $f$ indicates the density of field lines with respect to the magnetic moment and $\rho$ is the angle between the emission points and magnetic moment. The emission follows a power law $l^q$ expression for emission from a radial distance $l$ from the center of the star. For curvature-radiation emitting radio waves, the parameter $q$ is taken to be $-0.5$ \citep{ruder}. The normalized intensity $I_{norm}$ has been plotted in figure \ref{int} as a function of the misalignment angle $\chi$. The initial angle of misalignment is taken to be initially coinciding with the line of sight. As the misalignment angle evolves, the intensity falls sharply and then stabilizes to some final value of intensity which is orders of magnitude lower than the peak intensity. Thus after the PT, a pulsar may disappear from view due to much decrease in intensity. However, the opposite may also happen, and pulsars that were not visible before could evolve and cross the line of sight due to an increase in intensity and may become visible.\\

	\section{Summary \& Conclusion}
	
	The PT from NM to QM in NSs is a dynamic process that involves shock waves converting matter from nuclear to 2f followed by weak decay from 2f to 3f. As the combustion front evolves with time, the neutrinos are emitted as a function of time. Depending on the star's temperature and EoS, the neutrino-antineutrino annihilation energy deposited on the surface of the star has been calculated as the PT takes place inside the star. As the star loses energy due to PT, the star undergoes physical changes such as misalignment evolution and change in rotational velocities, calculated using the free precision model. Thus the phenomena of PT in NS can be captured by studying the multimessenger signal coming from NS like the neutrino energy, evolution of the star tilt axis, and detecting GWs.
	
	Due to the generation of neutrinos during the PT process's weak combustion, a massive amount of energy is deposited at the star's surface. The total energy deposited by the star is around $10^{49}-10^{50}$ ergs at timescales of the order of milliseconds. 
	Due to the energy loss, the star's tilt angle can evolve up to $\pi/15\;(12^o)$, thus evolving towards alignment. The misalignment angle changes are reflected in the continuous emission of gravitational waves whose amplitude is of the order of $10^{-27}-10^{-31}$. The amplitude and frequency of the gravitational wave signature change after PT due to change in misalignment angle and rotational frequency. The actual process of PT has much stronger signals (amplitude is of the order of $10^{-20}$) and last at max for a few ms.\\
	
	The energy and time signatures for our model fall well within the observed short gamma-ray burst signatures. The isotropic energy emitted on the rest frame of the source $E_{iso}$ is observed from the earth surface, which could come from PT in NSs. 
	Although the continuous GW signals are way beyond the present operating detectors' capabilities, the strong GW burst signal from the PT process itself is well within detection capability.
	A GW emission from the direction of observed SGRB's could show PT's signatures and would indicate QS to be one of the sources of SGRBs. The sudden disappearance or appearance of radio emissions from the particular pulsars emitting GRS bursts and showing short bursts of GWs could also indicate a PT. 
	
	Other observational signatures of misalignment evolution could be changing in the intensity of pulses from magnetic field emission of pulsars in addition to its standard spin down. Time signatures would uniquely differentiate the intensity variation due to spin down compared to intensity variation due to PT and can be investigated in the future. The short time signatures can also indicate a connection between fast radio bursts and gamma-ray bursts, and its possible common source could be a further extension of the present study.
	
	\section{Acknowledgement}
	DK thanks CSIR, Govt. of India for financial support. RM is grateful to the SERB, Govt. of India for monetary support in the form of Ramanujan Fellowship (SB/S2/RJN-061/2015). RP would like to acknowledge the financial support in the form of INSPIRE fellowship provided by DST, India. SS, RM, RP, and DK would also like to thank IISER Bhopal for providing all the research and infrastructure facilities.

\end{document}